\input{aipcheck}

\documentclass[
    ,final            
  ]
  {aipproc}

\layoutstyle{6x9}

\begin{document}

\title 
      [The hunt for cosmic neutrino sources]
      {The hunt for cosmic neutrino sources with IceCube}


\author{Elisa Bernardini$^{1}$ \newline (on behalf of the IceCube Collaboration)}{
  address={$^1$DESY, Platanen Allee 6, 15738 Zeuthen, Germany \\ corresponding author: elisa.bernardini@desy.de},
  email={elisa.bernardini@desy.de}
}

\begin{abstract}
IceCube is a cubic-kilometer neutrino telescope under construction at the geographic South Pole. Once completed it will comprise 4800 optical sensors deployed on 80 vertical strings at depths in the ice between 1450 and 2450 meters. Part of the array is already operational and data was recorded in the configurations with 9 (year 2006/2007), 22 (year 2007/2008) and 40-strings (year 2008/2009) respectively. Here we report 
preliminary results on the search for point-like neutrino sources using data collected with the first 22 strings (IC-22). 
\end{abstract}

\maketitle

\section{
Rationale of High Energy Neutrino Astronomy}
High energy neutrino telescopes attempt to resolve the origin of cosmic rays by detecting neutrinos. Whatever the sources of cosmic rays are, they are expected to emit high energy neutrinos as well. 
While hard to detect, neutrinos have the advantage of representing unique fingerprints of hadron interactions and, therefore, of the sources of cosmic rays. Moreover, their directions are not scrambled by magnetic fields. 

The candidates targeted by high-energy (E$>$1 TeV) neutrino telescopes are astrophysical objects showing clear evidence of non thermal emission and providing hints for cosmic ray acceleration.
A class of most favored candidates for cosmic ray acceleration up to the "Knee" are supernova remnants (SNRs), while only two types of objects are considered capable of accelerating protons at higher energies (up to $10^{20}$ eV): gamma-ray-bursts (GRBs) and active-galactic-nuclei (AGNs). 
These offer  the brightest events in the universe but mostly lie at cosmological distances, hard to observe with both protons and photons.
While high energy cosmic rays interact with the cosmic microwave background and photons with the extragalactic background light  (currently the maximum distance tested with observed high energy gamma-rays is z=0.536~\cite{MAGIC:3c279}), neutrinos would allow to extend the observable horizon of the universe. 

High energy gamma-rays are successfully detected for two decades from a large variety of sources (galactic and extragalactic) using satellites and air shower detectors.
Moreover, the energy spectra can now be investigated in detail by new generation Cherenkov telescopes, like H.E.S.S., MAGIC and VERITAS.
However, the observation of high energy gamma-rays alone does not provide unambiguous proof of hadron (i.e. cosmic rays) acceleration, since in most cases, the observed fluxes and spectra can be ascribed to high energy electrons and positrons which, once accelerated, give rise to gamma-rays either via inverse Compton scattering or bremsstrahlung.

In parallel Ultra high energy cosmic rays astronomy, is hunting for the sources of cosmic rays at energies larger than about 10$^{19}$ eV, where the gyroradius of a proton in the galactic magnetic field is larger than the thickness of the galactic disk and the trajectory deflections are reduced. However, the cosmic ray flux at these energies is so small that the statistics so far collected does not allow us to pinpoint their origin. The Auger detector is designed to collect sufficient statistics to address this question and its recent results provided evidence of anisotropy in the arrival directions of the highest energy cosmic rays (at energies larger than about 60 EeV~\cite{auger:science07}). However, the possibility of identifying individual sources is debated (see for example~\cite{icrc2007:dermer, Tinyakov:UHEcosmicrays}).

Several cases of high energy gamma-rays from SNRs have been recently discovered (e.g.~\cite{Voelk:VHE}). A remarkable example is RX J1713.7-3946, for which 
observational data favor the hypothesis of cosmic ray (proton) acceleration (e.g. the hadronic model in~\cite{voelk:RXJ1713} as opposed to the leptonic model in~\cite{HESS:RXJ1713}). 
X-ray measurements by Chandra, combined with detection of gamma-rays of energies up to at least 100 TeV, provide indeed a strong indication for acceleration of protons to energies of 1 PeV and beyond~\cite{Aharonian:RXJ1713cosmic}. This source and similar ones may give measurable fluxes of high-energy neutrinos as well, if  gamma-rays are generated by the decay of neutral pions (see for example~\cite{kappes:galactic,vissani:SNR} for neutrino fluxes from SNRs and~\cite{beacom:Cygnus} from the Cygnus region). Binary systems are also  promising sources of neutrinos (e.g.~\cite{aharonian:LS5039} for LS 5039), but various scenarios are predicted for the observed high-energy gamma-ray emission, each with different implications for the associated neutrino signal (e.g.~\cite{TorresHalzen2007}). Other objects also provide a good target for neutrino observations, whenever the observed gamma-rays have energies exceeding tens of TeV, like MGRO~J1908+06, observed by H.E.S.S. up to about 40 TeV~\cite{HESS:J1908} and by Milagro up to about 100 TeV~\cite{MILAGRO:J1908}. The associated neutrino fluxes may be detectable by IceCube (e.g.~\cite{kappes:MGROJ1908}).

Among extragalactic sources, an intriguing case is the blazar 1ES 1959+650, which manifested in 2002  an exceptional enhancement in its high energy gamma-ray emission which was not accompanied by an increase in the X-ray fingerprint of the entity expected in most leptonic scenarios~\cite{1ES1959:orphan}. This so called "orphan flare" is considered an hadronic indicator. AMANDA reported high energy neutrinos in possible temporal coincidence with this period of exceptional activity, but this result is not statistically significant~\cite{icrc2005ICECUBE}. Hadronic  interpretations of this flaring episode also did not yield clear conclusions on the genuine nature of the detected neutrinos as a possible cosmic signal~\cite{reimer2005,halzen2005}. In order to reach the sensitivity required by most astrophysical scenarios to detect cosmic neutrinos, at least cubic-kilometer-scale arrays are  required~\cite{halzen:science2007}. For models predicting high energy neutrinos in AGNs see for example~\cite{muecke:protonblazar} (and references therein), and~\cite{waxman, waxmanbahcall} for neutrinos from GRBs.

Large data samples from AMANDA-II~\cite{andres2000} have been analyzed, showing no evidence of point-like signals from the northern sky in either time-integrated or time-variable searches~\cite{achterberg:5yrpointsources,icrc2005ICECUBE,icrc2007ICECUBE}. In~\cite{achterberg:5yrpointsources}, the sensitivity to tau neutrinos was estimated for the first time. An improved analysis based on a maximum likelihood method was later developed and an analysis of the complete AMANDA data set (about 3.8 years of effective lifetime) has been presented, yielding the most stringent flux upper limits to point sources of neutrinos~\cite{icrc2007ICECUBE}.
Results based on IceCube data (9 strings) have also been shown. No indication of point sources has been found yet~\cite{icrc2007ICECUBE}. 

Neutrinos in association with GRBs are also being searched for in dedicated analyses. Besides promising sources of neutrinos, they offer the advantage of higher signal-to-noise ratio compared to steady point-sources due to the very small burst time window in which the neutrino signal is expected. Since IceCube has a large field-of-view, a very large portion of the sky is simultaneously observable and data can be scanned to look for correlations with GRBs catalogs. The best limits so far obtained are from AMANDA data. Optimistic predictions are ruled out~\cite{achterberg:2008GRB}.

New search strategies are recently being developed, aiming at increasing the chance to discover comic neutrinos by on-line searches for correlations with established signals (e.g. flares in high-energy gamma-rays) triggered by neutrino observations (neutrino-triggered target-of-opportunity). For sources which manifest large time variations in the emitted radiation, the signal-to-noise ratio can be increased  by limiting the neutrino exposures to most favorable periods. The chance of discovery can then be enhanced (the so-called "multi-messenger approach") by ensuring a good coverage of simultaneous data at a monitoring waveband (e.g. gamma-rays). The first realization of such an approach led to two months of follow-up observations of AMANDA triggers by MAGIC~\cite{icrc2007ps:NToO}. An extension of this program to IceCube and also to optical follow-up observations is in progress.

A search for a cosmic neutrino signal is finally performed by looking for distortions in the background properties at high energies (above TeV), where the flux of atmospheric neutrinos is much reduced. Also in this case only upper limits are so far derived  (cfr. ~\cite{achterberg:4yrdiffuse} for limits on diffuse fluxes from unresolved sources across the northern sky based on AMANDA data in the energy range of 16 TeV to 2.5 PeV,~\cite{ackermann:UHE} for limits across the horizon in the energy range of 0.2 PeV to 1 EeV and for all-sky limits from cascade-like events~\cite{icrc2007ICECUBE}).  Results from IceCube (nine strings) have been reported in~\cite{icrc2007ICECUBE}.

Besides hunting for astrophysical neutrino sources, IceCube is dedicated to particle physics topics including the indirect search for Dark Matter, the study of Lorentz Invariance, the study of neutrino oscillations and the search for magnetic monopoles.

Here we report the latest results on the search for neutrinos from point-like sources, as obtained from the analysis of data collected with IceCube (22-strings configuration), with an effective lifetime of 276 days.

\section{The IceCube telescope \& operation principle}
IceCube is a cubic-kilometer neutrino telescope under construction at the South Pole. Once completed in 2011, it will comprise 4800 Digital Optical Modules (DOM) along 80 vertical strings deployed in the ice at depths from 1450 to 2450 meters. The first nine in-ice strings (IC-9) recorded data in 2006-2007~\cite{achterberg:ic9} and the first 22 in-ice strings (IC-22) from May 2007 to April 2008. Forty strings (half array) are operational since May 2008. 

The neutrino telescope is complemented with an air-shower array (IceTop)  which covers a surface of one square kilometer and consists of 80 detector stations, each equipped with two ice Cherenkov tanks. IceTop serves as a veto to cosmic ray showers and as a calibration array, and covers in addition science topics on cosmic rays physics. In 2006, the particle spectrum of the solar flare of December 13 was measured in the energy range between 0.6 and 7.6 GeV~\cite{IceTop:solarflare2006}.

IceCube is optimized to detect  neutrinos of all flavors in the energy range from TeV to EeV. Signal waveforms from individual photomultipliers are digitized in each DOM and transmitted to the surface by twisted-pair cables. Data is processed by several hardware and software triggers, responsive to different event topologies. 
A series of reconstruction algorithms and event filters are applied in situ to calibrated data to reduce the trigger rate (about 650 Hz for the SMT with 22 IceCube strings) to a total data volume of about 30 GB/day, 
which was then transferred north via satellite. Events are then reconstructed using more CPU intensive maximum-likelihood techniques, in which events are fitted to templates representing different neutrino interaction modes. For example, $\nu_\mu$ and $\nu_\tau$ CC interactions can lead to muon tracks (in case of $\nu_\tau$ arising from the muonic decay mode of the tau lepton), which, at TeV energies, provide a good handle on the primary neutrino direction. On the other hand, $\nu_e$ and $\nu_\tau$ CC,  as well as all-flavor NC interactions, can lead to compact energy deposits (cascades, in case of $\nu_\tau$ CC interactions arising from both the hadronic and electronic decay modes of the tau lepton), which provide a good handle on the primary neutrino energy.

\section{Results on point-like sources of neutrinos with 22 IceCube strings}
Searches for astrophysical sources of neutrinos have to cope with background events
from the interaction of cosmic rays in the Earth's atmosphere. Decays of secondary mesons
induce downward going muons and a more isotropic flux of neutrinos.
Rates measured with 22 IceCube strings are $O(10^9)$ events/year from downward-going atmospheric muons and $O(10^3)$ events/year from atmospheric neutrinos. In view of point-source searches, neutrino
candidates are typically selected starting with angular cuts, i.e. rejecting downward-going muon tracks, since
only neutrinos can travel through the Earth. This limits the sensitivity to the northern sky. An effort is on-going to overcome this limit and extend both the field of view and the energy window of sensitivity (i.e. multi-PeV)~\cite{icrc2007ICECUBE}.

Three independent analyses developed for point source searches with 22 IceCube strings in the energy region from TeV to PeV are presented here: a binned search, based on an event counting approach using circular angular search bins, and two un-binned searches, one applied to IceCube data and based on a maximum likelihood approach which models the signal expectation based on the detector point spread function and a simple energy estimator,  and another  optimized for soft spectra, applied to combined IceCube and AMANDA events. 

The event selection is based on similar parameters in all three analyses. First, a filter based on event direction and quality parameters is used to reduce atmospheric muons by about three orders of magnitude. Yet an irreducible background remains from upward-going muons induced by atmospheric neutrinos together with a small fraction of mis-reconstructed downward-going muons plus possible signal events. A signal from cosmic neutrinos would manifest itself as an excess of events over the background at a specific location in the sky. Further track quality cuts are therefore applied, designed to achieve a good angular resolution, including angular uncertainty estimates, the number of un-scattered photon hits with a small time residual w.r.t.
the Cherenkov cone, a likelihood ratio comparing a prior of an atmospheric muon track (downward-going) to the reconstructed one (upward-going), the reduced likelihood of the directional reconstruction,
etc. 
To avoid biases in the event selection, the final cuts are optimized in a blind approach. In the search for point sources, where the event direction is used to look for a signal, this is accomplished by fixing cuts on a sample of events with randomized right ascension. In a given declination band (of size comparable to the angular resolution), a negligible point source neutrino contribution compared to the total background is expected, allowing an estimate of the background from scrambled data. Details of each analysis and respective results are given below.

\begin{description}
\item[The IC22 binned analysis] distinguishes a localized excess of signal from a uniform background using circular angular search bins. The sensitivity (average upper limit in case of no signal), in a given declination band, can be obtained from Poisson statistics and it is calculated by comparing the
number of simulated signal events for a given flux $\Phi^0$ reconstructed
inside the search bin, $n_{sig}$, to the average number
of background events as measured from scrambled data $n_{bg}$, using~\cite{achterberg:5yrpointsources}:
\begin{equation}
\Phi^{lim}=\Phi^0\cdot \mu_{90}(n_{bg})/n_{sig}(\Phi^0)
\end{equation}
where $\mu_{90}$ is the event upper limit at 90\% CL obtained in the absence of signal following the Feldman\&Cousins ordering principle~\cite{feldman1998}.
Cuts on track quality parameters were optimized considering the detection and reconstruction efficiencies as a function of declination. The search bin radius was optimized as an additional free parameter (mean value = 2.1$^\circ$). Final cuts were chosen to deliver an optimum sensitivity for a compromise between a spectral index $\gamma=2$ and $\gamma=3$ for the neutrino energy spectra.
This analysis uses a different event directional reconstruction then the other two analyses, that delivers better angular resolution at very high energies. 
However, it uses reduced event information (events are equally weighted, disregarding the individual directional error and energy) and it is therefore expected to achieve a lower sensitivity. 2956 events have been selected. 
The atmospheric neutrino content of the samples
is greater than 90\%. The rest comes from mis-reconstructed down-going muons, mostly multiple events from independent cosmic ray showers. From simulation, a sky- and energy-averaged median angular resolution of 1.3$^\circ$ is estimated for signal neutrinos with spectral index $\gamma=2$. 

\item[The IC22 un-binned analysis]
constructs a likelihood function which depends
on the signal probability density function (pdf), obtained from MC simulation of the event energy and angular distribution, the background pdf (obtained from data as a function of the event position $x_i$), and total number of data events $n_{tot}$~\cite{braun:unbinned}:
\begin{equation}
L = \prod_{i=1}^{i=n_{tot}} \frac{n_s}{n_{tot}} S_i(x_i, x_s,E_i, \gamma) + \frac{n_{bg}}{n_{tot}}
B_i(x_i,E_i).
\end{equation}
The signal pdf $S_i$ is characterized by each event's direction, $x_i$, angular uncertainty
around $x_s$, energy estimation, $E_i$ (i.e. the number of DOMs with at least one hit), and 
the assumed energy spectrum. The number of signal events, $n_s$, is found by maximizing
the likelihood ratio of the background plus signal hypothesis against the background-only case.
5114  events have been selected. From simulation, a sky- and energy-averaged median angular resolution of 1.5$^\circ$ is estimated for signal neutrinos with spectral index $\gamma=2$. Based on signal simulation with a spectral index $\gamma=2$ this analysis provided both the best discovery and limit setting capability and was therefore a-priori chosen to deliver the final significances and upper limits.  The enhancement gained by using the un-binned method (without the systematic  error) is 35\%. 
It also shows on average a factor of 2 improvement over that of the total statistics collected by AMANDA-II~\cite{icrc2007ICECUBE}
and provides the best limits to date. 
\end{description}
\begin{figure}	
 \includegraphics[height=.27\textheight]{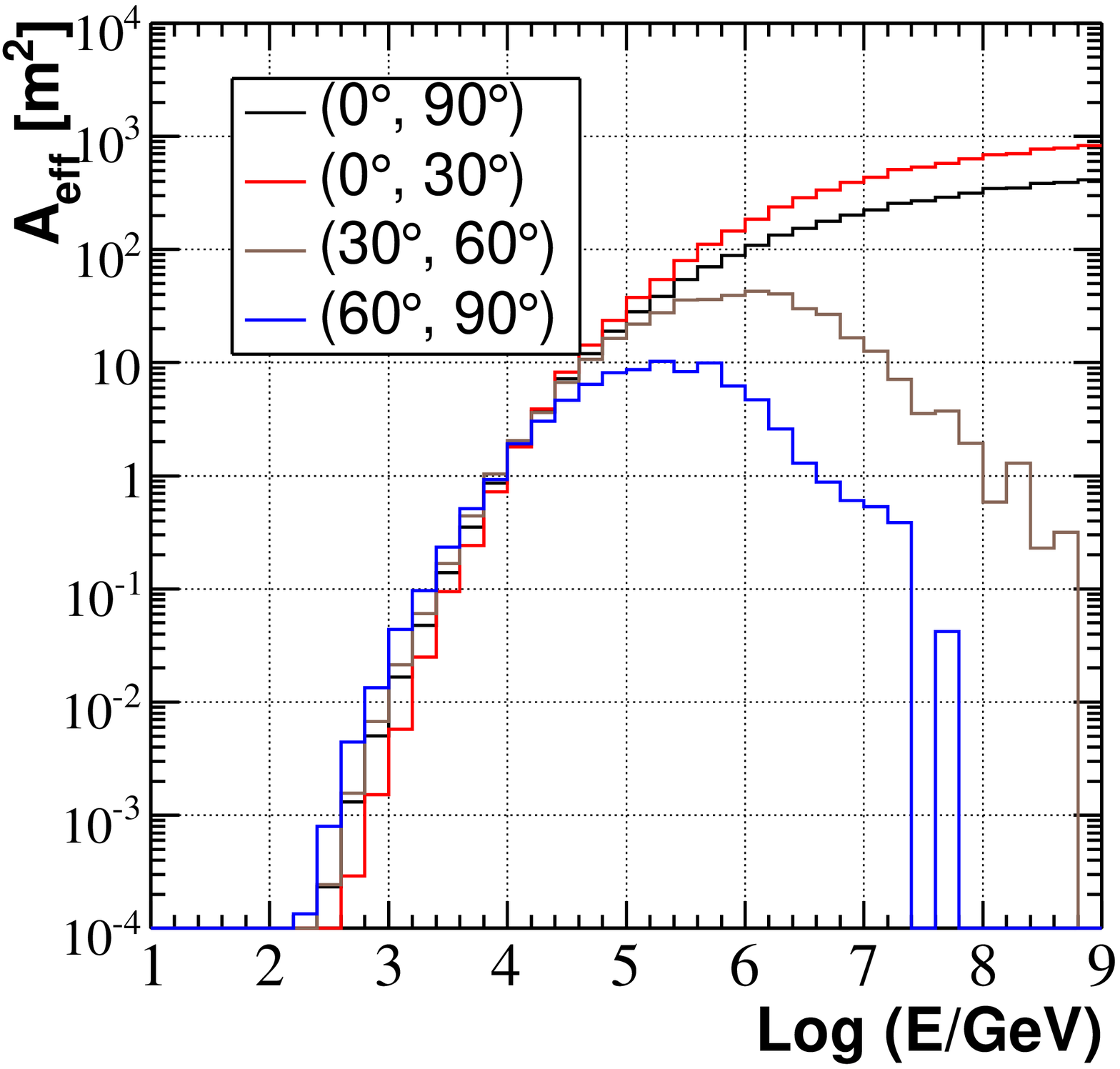}
 \includegraphics[height=.27\textheight]{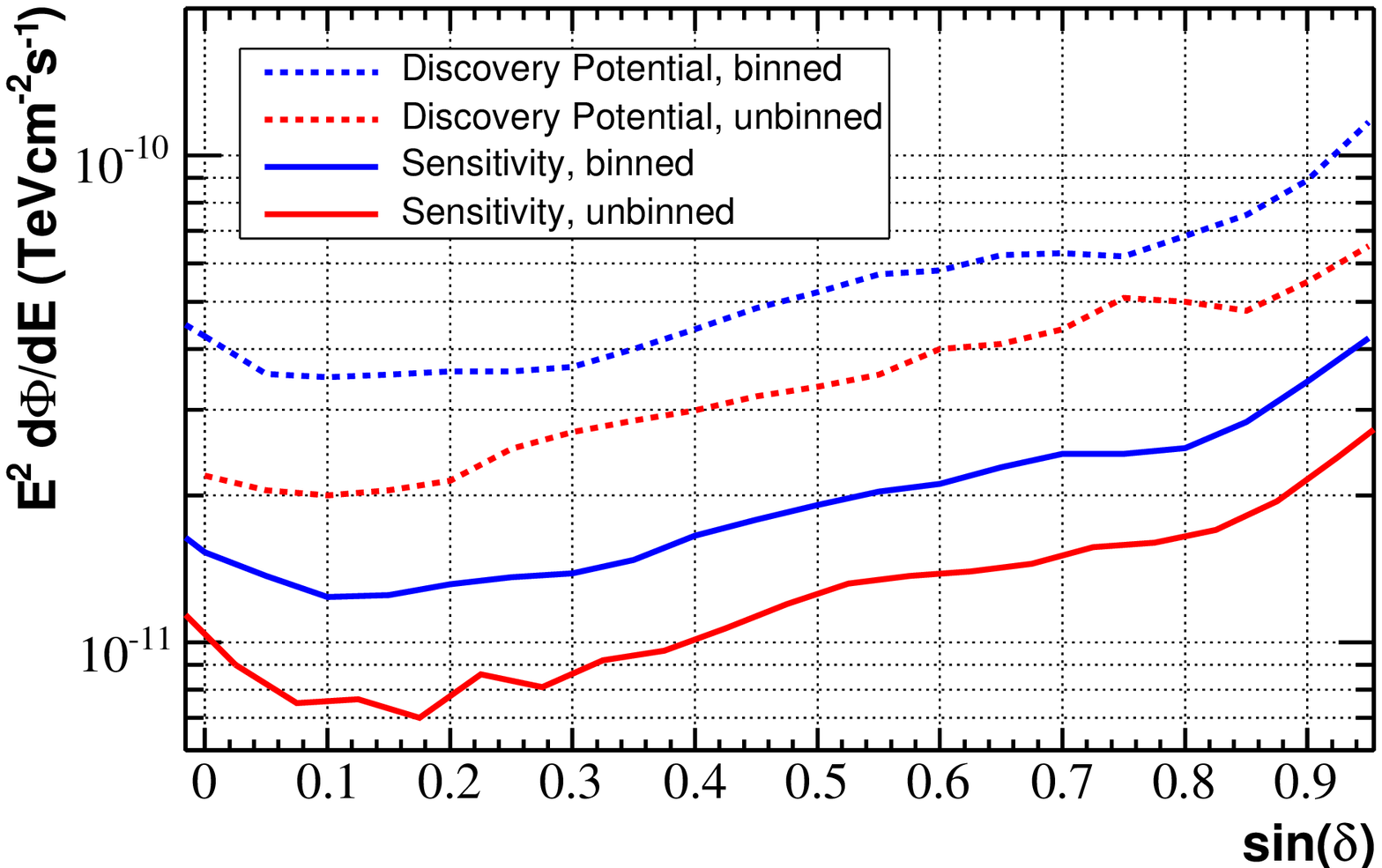}
  \caption{{\bf Left:} Effective area of IceCube with 22 strings versus energy and declination. The decrease at high energies for vertical tracks is due to absorption of neutrinos in the Earth. {\bf Right:} Sensitivity (average upper limit in case of no signal, at 90\% CL) and discovery potential (minimum flux delivering a 5$\sigma$ detection with 50\% probability) for the binned (blue) and un-binned (red) analysis. The latter shows better performance, due to event weighting including the angular reconstruction error and a simple energy estimator.}
  \label{effarea}
\end{figure}
The final effective area of the binned analysis is presented in Fig.~\ref{effarea} and is similar for the two analyses (binned and un-binned). Sensitivities and discovery potentials, as functions of declination, are also compared in Fig.~\ref{effarea}. The
sky-average sensitivity at 90\% 
CL to a generic flux (with spectral index $\gamma=2$) is $\Phi_{\nu_\mu}\cdot$~E$^2\leq$1.3 (2.0) $\cdot 10^{-11}$~TeV~cm$^{-2}$s$^{-1}$  for the un-binned (binned) method respectively. The statistical overlap of the event samples tested by the two analyses is 72\%.

A pre-defined list of 28 sources (galactic and extragalactic) gave no significant excess over the expected background.
The biggest deviation was found for 1ES 1959+650 (Crab Nebula) with a post-trial p-value
of 66\% (91\%) by the un-binned (binned) analysis. Details are given in Tab.~\ref{catalog}.

\begin{table}
\begin{tabular}{lcc} 
\hline
  & \tablehead{1}{r}{b}{binned}
  & \tablehead{1}{r}{b}{un-binned\tablenote{Negative excesses not treated by the un-binned analysis.}}\\
\hline
MGRO J2019+37 & 0.51 & 0.25 \\
MGRO J1908+06 & 0.90 &      $^*$\\
TeV J2032+4130 & 0.81 &      $^*$\\
SS 433                   & 0.66 & 0.32 \\
Cyg X-1                & 1.00 &      $^*$\\
LSI +61 303          & 0.80 & $^*$ \\
GRS 1915+105   & 0.20 &       $^*$\\
XTE J1118+480  & 1.00 & 0.08  \\
GRO J0422+32   & 0.15 &       $^*$\\
Geminga             & 0.51 &       $^*$\\
Crab Nebula   & {\bf 0.10} &     $^*$\\
Cas A                   & 0.54  &     $^*$\\
Mrk 421              & 0.82  &     $^*$\\
Mrk 501              & 0.48  &     $^*$\\
1ES 1959+650   & 0.57  & {\bf 0.07} \\
1ES 2344+514   & 0.19  &      $^*$\\
H 1426+428        & 1.00   &      $^*$\\
1ES 0229+200   & 0.81   &      $^*$\\
Bl Lac                & 0.80   & 0.37  \\
S5 0716+71        & 0.62   & 0.31  \\
3C66A                  & 0.77 & 0.31   \\
3C454.3               & 0.13 & $^*$\   \\
4C38.41               & 0.51 &        $^*$\\
PKS 0528+134   & 1.00 &        $^*$\\
3C 273                 & 0.88 & 0.37   \\
M87                      & 0.68 &         $^*$\\
NGC 1275           & 0.49 & 0.21    \\
Cyg A                  & 0.19 &  $^*$ \\
\hline
\end{tabular}
\caption{Pre-trial p-values from the binned and un-binned analysis of a pre-defined catalog of 28 sources (preliminary). Sources were selected based on evidence for non-thermal emission. In bold  are marked the respective highest excesses.}
\label{catalog}
\end{table}

\begin{figure}
  \includegraphics[height=.23\textheight]{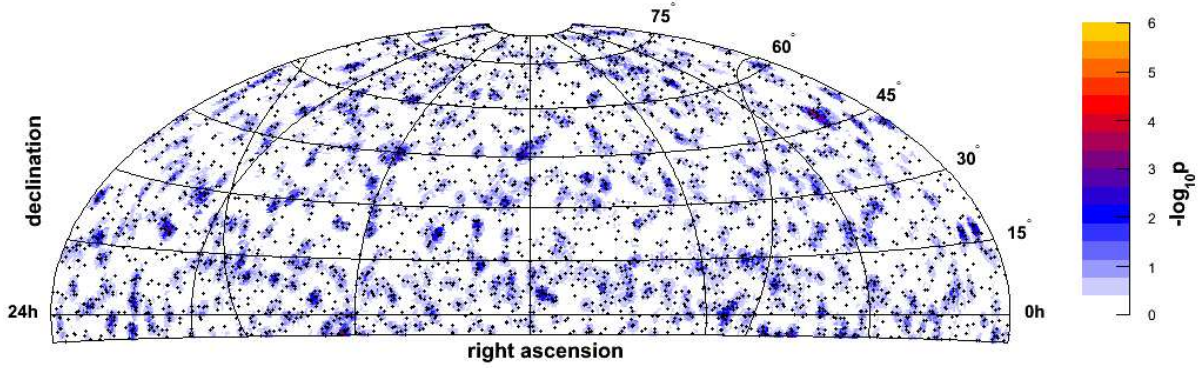}
  \caption{Pre-trial significance sky map from the binned analysis (preliminary).}
  \label{skymap_binned}
\end{figure}

\begin{figure}
  \includegraphics[height=.23\textheight]{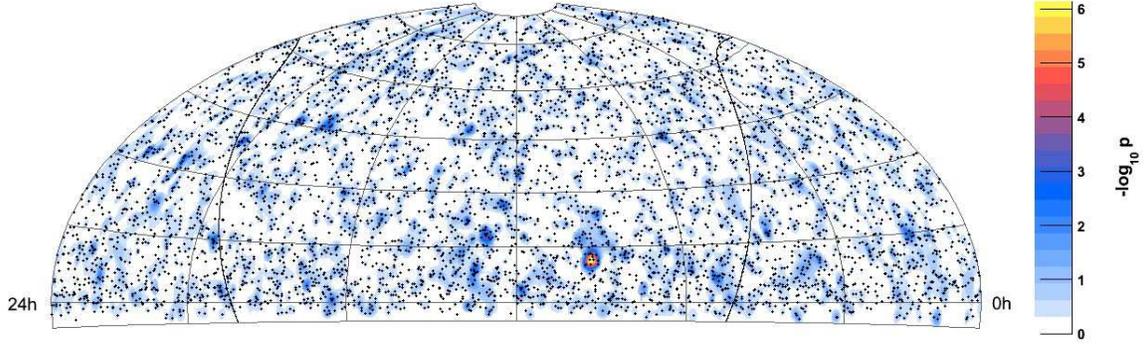}
  \caption{Pre-trial significance sky map obtained with the un-binned analysis (preliminary).}
  \label{skymap_unbinned}
\end{figure}

The results from the all-sky search for both analyses are  shown in Fig.~\ref{skymap_binned} and Fig.~\ref{skymap_unbinned}. The maximum
positive deviation from the background hypothesis was found at right ascension $\alpha=$153$^\circ$ (183.5$^\circ$), declination $\delta= 11^\circ$ (47$^\circ$), with a post-trial p-value of 0.67\% (67.6\%) for the un-binned (binned) analysis respectively. 
The post-trial p-value for the highest excess is 1.3\%, taking into account a trial factor of two arising from the multiple tests performed (i.e. pre-selected catalog and northern sky un-biased search). The spot found by the un-binned analysis (a-priori chosen to deliver the final significances) was further investigated (e.g. by a time-dependent analysis) resulting in no significant event cluster in time (no events are closer together than 10 days). It will be tested with better sensitivity using data with 40 strings. The difference in the p-value found by the two analyses arises mainly from the energy parameter implemented in the likelihood function.

\begin{description}
\item[The IC22+AMANDA low-energy optimized analysis] is designed for galactic sources, which are expected to exhibit softer spectra than a simple power-law with $\gamma=2$. A likelihood function, similar to the one given above, is implemented:
\begin{equation}
L = \prod_{i=1}^{i=n_{tot}} \frac{n_s}{n_{tot}} S_i + (1-\frac{n_s}{n_{tot}}) B_i.
\end{equation}
with constant background pdf $B_i=1/\Omega$ and a signal pdf $S_i = \exp(-\psi^2/(2{\sigma_i}^2 ))/2\pi\cdot{\sigma_i}^2)$, with $\psi$ being the angle between source sky position and each event's reconstructed direction.
All events in a $\pm 5^\circ$ band around the galactic plane (covering a solid angle $\Omega$) are included, with a lifetime of 276 days. 8727 events have been selected. 
From simulation, a sky- and energy-averaged median angular resolution of 1.5$^\circ$ is estimated for signal neutrinos with spectral index $\gamma=2$. 

Four pre-selected sources (Crab Nebula, Cas A, SS433, LSI+61 303) were also tested with this method.
\end{description}
This  analysis also found no significant deviation from the background hypothesis, with the highest excess in the galactic scan found at longitude $l=75.9^\circ$, latitude $b=2.7^\circ$ (post-trial p-value 95\%) and the highest excess among the four pre-selected sources found in the direction of the Crab Nebula (post-trial p-value 37\%).

We conclude that no significant deviation from the background hypothesis was found in the IceCube data collected with 22 strings by the three analyses presented here. All results shown are preliminary. Flux upper limits, including the systematic uncertainties, are being calculated.

\section{Summary \& Outlook}
The IceCube neutrino telescope is now more than half-way completed with 40 strings in operation in 2008. We have shown preliminary results on the search for point-like signals in the northern sky, using data collected from May 2007 to April 2008 with 22 strings. No significant excess of events was found by three independent analyses. An additional work, focused on extending the field of view to the southern sky for multi-PeV neutrinos, will be presented elsewhere soon. Analysis of data in the 40 strings configuration is on-going. 

Data collected with 22 strings is also being searched for an overall excess due to un-resolved sources (diffuse flux). This work is 
expected to exceed the current best limits obtained with AMANDA-II (4 years of data), due to improved energy estimation and the larger acceptance above 100 TeV. 

The full IceCube neutrino telescope is expected to  achieve a track angular resolution of less than one degree and an order of 
magnitude improvement over the sensitivity to point-like neutrino sources obtained with the entire data set collected with AMANDA-II (7 years of data), within three 
years of operation. Based on current estimations, the cubic-kilometer scale neutrino telescope has good chances  to identify fluxes from steady point sources of high energy neutrinos.   One year of data with 80 strings is expected to achieve a sensitivity to diffuse fluxes of neutrinos currently predicted from unresolved AGNs. 

Detection of high energy neutrinos with the next generation neutrino telescopes (IceCube, KM3NeT) will probe the most powerful accelerators and the physical mechanism leading to their non-thermal emission. Increased discovery potential and a more comprehensive phenomenological understanding is expected by combining neutrino data with established non-thermal astrophysical signals (e.g. high energy gamma-rays in the so-called "multi-messenger approach").

\begin{theacknowledgments}
We acknowledge the support from the following agencies: U.S. National Science 
Foundation Office of Polar Programs, U.S. National Science Foundation Physics Division, University of Wisconsin Alumni Research Foundation, U.S. Department 
of Energy and National Energy Research Scientific Computing Center, Louisiana 
Optical Network Initiative (LONI) grid computing resources, Swedish Research
Council, Swedish Polar Research Secretariat, Knut and Alice Wallenberg Foundation (Sweden), German Ministry for Education and Research (BMBF), Deutsche 
Forschungsgemeinschaft (DFG), (Germany), Fund for ScientiÞc Research (FNRS-FWO), Flanders Institute to encourage scientiÞc and technological research in industry (IWT), Belgian Federal Science Policy Office (Belspo), and the Netherlands 
Organisation for ScientiÞc Research (NWO); M. Ribordy acknowledges the support 
of the SNF (Switzerland); A. Kappes and A. Gross acknowledge support by the EU 
Marie Curie OIF Program; M. Stamatikos is supported by an NPP Fellowship at 
NASAGSFC administered by ORAU. 
\end{theacknowledgments}

\bibliographystyle{aipproc} 
\bibliography{EBernardini}

\end{document}